\documentclass[%
 reprint,
 amsmath,amssymb,
 aps,
]{revtex4-2}

\usepackage{graphicx}
\usepackage{dcolumn}
\usepackage{bm}

\newcommand{\orcid}[1]{\hspace{1mm}\href{https://orcid.org/#1}{\includegraphics[height=0.3cm,keepaspectratio]{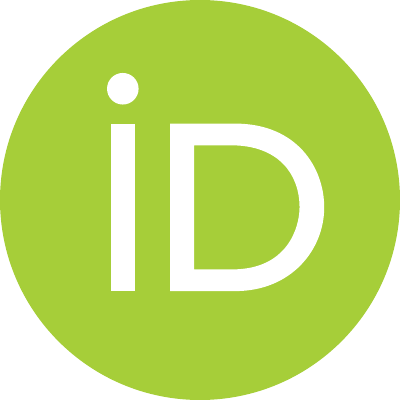}}}

\newcommand{\aap}{A\&A}

\newcommand{\apjs}{ApJS}
\newcommand{\mnras}{MNRAS}

\newcommand{\physrep}{Phys. Rep.}

\newcommand{\jcap}{JCAP}

\newcommand{\aapr}{AApR}
\newcommand\apjl{{ ApJL}}%

\begin{document}

\preprint{APS/123-QED}

\title{The impact of the eROSITA bubbles on Galactic cosmic-ray transport}

\author{Benedikt Schroer\orcid{0000-0002-4273-9896}}
\email{bschroer@uchicago.edu}
\affiliation{Department of Astronomy and Astrophysics, University of Chicago, 5640 S Ellis Ave, Chicago, IL 60637, USA}

\begin{abstract}
We propose that the observed spectral hardening in Galactic cosmic ray fluxes is governed by macroscopic Galactic outflows, such as the eROSITA bubbles, rather than microphysical variations in their scattering properties. Employing a phenomenological transport model, we show that an advective outflow boundary naturally reproduces the $300\,$GV hardening in secondary-to-primary ratios. Global fits to precision AMS-02 data yield an effective local inner halo boundary of $\sim 5\,$kpc and an outflow speed of $\sim 360\,$km/s, in striking agreement with independent multi-wavelength kinematic constraints of the eROSITA outflows. This interpretation provides a testable alternative to breaks in the effective diffusion coefficient, without increasing the number of free parameters.
\end{abstract}

\maketitle

\section{Introduction}
\label{sec:intro}

Experiments such as AMS-02 \cite{ams21}, DAMPE \cite{dampe21} and CALET \cite{CALET19} provide cosmic-ray (CR) nuclei flux data of several elements with percent-level precision.
This unprecedented accuracy revealed unexpected features in their fluxes, such as a hardening around $300\,$GV in nuclei fluxes and secondary-to-primary ratios \cite{ams21,pamela11, dampe22,calet22} and a softening around $10\,$TV in fluxes of several elements \cite{dampe21,dampe25}.
Despite this, measurements are well described by a steady-state halo model, in which CRs diffuse in a magnetized halo $3-7\,$kpc in size \cite{evoli+19b,schroer+21b,maurin+22,chernyshov+24}.
To incorporate these features, phenomenological models introduce ad-hoc breaks in the diffusion coefficient \cite{genolini+17,evoli+19b,schroer+21b,boschini+20,dimauro+24,delatorreluque+22} and a maximum energy of the majority of sources around $10\,$TV \cite{recchia+24}.

How such a halo forms and the origin of such a break are often related to microphysics \cite{chernyshov+22,chernyshov+24,evoli+18b,blasi+12,ewart+26,recchia+24}.
Existing theories explain both phenomena by an interplay of CR self-confinement and extrinsic turbulence \cite{evoli+18b,blasi+12,aloisio+13}.
However, recent plasma simulations cast doubt on this explanation for extrinsic Alfvénic turbulence \cite{schroer+25a,schroer+25b}.
Other self-confinement models explain the halo as purely self-generated \cite{chernyshov+22,chernyshov+24}.
Under certain assumptions, self-confinement becomes more effective far away from the Galactic disk.
As a result, CRs couple strongly to Alfvén waves and are advected outwards with the local Alfvén speed at a given distance from the disk \cite{chernyshov+22,chernyshov+24}.
That distance defines the effective halo size and is energy dependent, extending up to $\sim 9\,$kpc at TV rigidities.
While relying on the properties of the background gas and magnetic field as a function of distance from the disk, these models explain both spectral features as pure transport phenomena.
Alternative explanations of the hardening invoke inhomogeneous scattering properties of CRs in the Galaxy \cite{tomassetti12,cowsik+14,cowsik+16,recchia+24,kempski+22,ewart+26}.
However, these theories remain observationally elusive, as inferring local inhomogeneous CR transport properties is currently impossible outside localized source regions (e.g., TeV halos \cite{hawc17_coll,LHAASO21}).

The aforementioned models primarily focus on microphysics or localized spatial variations, generally neglecting observational evidence for macroscopic Galactic outflows, in which CRs could be advected.
The X-ray detection of the massive eROSITA bubbles confirms the existence of such large-scale outflows extending from the Galactic Center \cite{predehl+20}. Crucially, multi-wavelength observations provide kinematic constraints on their expansion. 
UV absorption-line spectroscopy and radio emission of high-velocity clouds, combined with X-ray measurements of the forward shocks, indicate that the bubbles are driven by galactic winds with measurable outflow velocities (see e.g., \cite{mccluregriffiths+13,fox+15,bordoloi+17,diteodoro+18,lockman+20,ashley+20,sarkar24}).
These velocities range from a few hundred km/s for clouds in the outflow up to $\sim 1000\,$km/s near the base of the injection and at the shock fronts.
Common interpretations relate these structures to past Galactic Center activity, estimating them as persistent in time \cite{shimoda+24} or $\gtrsim 3-20\,$Myrs old \cite{bordoloi+17,predehl+20,yang+22,sarkar24}, exceeding the residence of CRs at $\sim 100\,$GV.
Furthermore, recent models place their closest edge at only $\sim 4\,$kpc away from the Sun \cite{liu+24}, surprisingly close to estimates of the diffusive halo size \cite{maurin+22,evoli+19b}.

Motivated by these considerations, we present a phenomenological model incorporating the eROSITA bubbles.
The onset of the outflow serves as a natural spatial limit for the local diffusive region.
In this picture, the derived transport parameters, such as the outflow velocity and halo size, are no longer solely constrained by CR data. Instead, they can be compared against existing multi-wavelength observations. 
Unlike previous Galactic wind models \cite{recchia+16,recchia+17}, the hardening here emerges from the distant, observed outflow rather than near-disk properties.

\section{Phenomenological Model}
\label{sec:model}
We construct a phenomenological model, capturing the essential physics. First, we treat the eROSITA bubble as a macroscopic outflow with a constant speed $u_2$ initiating at a fixed distance $H_1$ from the disk. Although the problem is inherently multi-dimensional, a 1D geometry is justified given the expected competing escape paths, as detailed in Section~\ref{sec:estimate}. 

Our advection speeds and diffusion coefficients are $u_1$, $D_1$ for $0 \leq |z| < H_1$ and $u_2$, $D_2$ for $H_1 \leq |z| \leq H_2$.
Magnetic turbulence inside the hot gas outflows could differ from the quiescent halo, motivating different diffusion coefficients.
With these assumptions, we solve the Galactic CR transport equation \cite{evoli+19b,schroer+21b}
\begin{multline}\label{eq:slab}
-\frac{\partial}{\partial z} \left[ D \frac{\partial I_a }{\partial z} \right]
+ \frac{\partial u I_a}{\partial z}
- \frac{du}{dz} \frac{1}{3} \frac{\partial I_a p}{\partial p}
\\
+ \frac{\partial}{\partial p} \left[ \left(\frac{dp}{dt}\right)_{a,\rm ion} I_a \right]
+ \frac{\mu v(p) \sigma_a}{m} \delta(z) I_a
+ \frac{I_a}{\hat\tau_{d,a}}
\\
= 2 h A p^2 q_{0,a}(p) \delta(z)
+ \sum_{a' > a} \frac{\mu\, v(p) \sigma_{a' \to a}}{m}\delta(z) I_{a'}
+ \sum_{a' > a} \frac{I_{a'}}{\hat\tau_{d,a'}}
\end{multline}
for the flux of CRs $I_a=Ap^2 f_a(p)$
where $A$ is the mass number and $f_a$ is the phase space density of isotope $a$. The first two terms on the left hand side describe diffusion and advection.
The third, fourth and fifth terms are adiabatic, ionization and spallation losses respectively, where the latter two are limited to the gaseous Galactic disk.
$\mu$ is the Galactic disk column density $\mu = mhn_0$, fixed according to Ref.~\cite{ferriere+01}, taking into account the interstellar-medium (ISM) composition and a disk half height $h$.
The last term on the left- and right-hand side describes radioactive decay of isotope $a$ or production of $a$ via the decay of $a'$.
The source terms on the right-hand side are injection for primaries in the disk with a power law in momentum $q_{0,a}(p) \propto q_a p^{-\gamma}$ and production of nucleus $a$ via the spallation of heavier elements $a'$.
We compute inelastic cross-sections using standard parametric fits \cite{tripathi+96,tripathi+97} and partial production cross-sections following Refs.~\cite{evoli+19a,evoli+18c}.

\subsection{Grammage}
\label{sec:grammage}
To understand how such a situation leads to a hardening, we consider the grammage accumulated during propagation by a stable nucleus diffusing in the inner halo.
As detailed in the Supplemental Material, the grammage can be approximated as
\begin{equation}
\begin{split}
    \chi & \approx \frac{\mu v}{2} \frac{H_1}{D_1} \left(1 + \frac{\frac{D_1}{H_1}}{u_2 + \frac{D_2}{(H_2-H_1)}} \right) \\
    &\approx 
    \begin{cases}
      \frac{\mu v}{2} \frac{H_1}{D_1}\,, & D_1/H_1 \ll u_2 \\
      \frac{\mu v}{2 u_2}\,,           & D_2/(H_2-H_1) \ll u_2 \ll D_1/H_1 \\
      \frac{\mu v}{2} \frac{(H_2-H_1)}{D_2}\,, & u_2 \ll D_2/(H_2-H_1)
    \end{cases}\,,
\end{split}
\end{equation}
where we illustrate three distinct physical regimes. In the low-energy limit $D_1/H_1 \ll u_2$, the expression reduces to $\chi \approx \frac{\mu v}{2} \frac{H_1}{D_1}$. This mirrors the standard diffusion-dominated halo model, where the halo size is defined by the distance to the outflow $H_1$. In this regime, CRs effectively free-escape once they reach the highly advective bubble boundary.

At intermediate rigidities, where $D_2/(H_2-H_1) \ll u_2 \ll D_1/H_1$, the grammage becomes independent of the diffusion coefficient: $\chi \approx \frac{\mu v}{2 u_2}$, matching the advection-dominated transport in the weighted slab model \cite{evoli+19a}. The transition between these first two limits demonstrates that a spectral hardening is naturally reproduced as the CR escape speed shifts from being dominated by diffusion over short distances to being limited by a constant advection outflow boundary.

Interestingly, a third regime could emerge in which $u_2 \ll D_2/(H_2-H_1)$. Here, diffusion in the extended outer zone dominates over advection with the outflow $\chi \approx \frac{\mu v}{2} \frac{(H_2-H_1)}{D_2}$. Consequently, this phenomenological model can, in principle, capture both a spectral hardening and a softening. In this highest-energy regime, the grammage and the rigidity of the softening depend only on the ratio $D_2/(H_2-H_1)$.
While a constant source grammage \cite{aloisio+13,ambrosone+25} could mask this softening in secondary-to-primary ratios, it would manifest in primary fluxes, providing an alternative to the assumption of a maximum rigidity cutoff at CR sources.

Stable CR nuclei fluxes are only sensitive to the ratio $D_2/(H_2-H_1)$. Contrary to lower energies, there are no measurements of unstable isotopes with decay times on the order of their residence time at these large energies. Hence, breaking the degeneracy with current data is impossible.
As a result, we use the ratio $D_2/(H_2-H_1)$ as a single fit parameter to fit the softening at $10\,$TV without loss of generality.
The diffusion coefficient in the inner region is parameterized as $D_1(R)=D_0 (R/R_0)^\delta$ with $R_0=10\,$GV.
Given the limited data points of DAMPE above the break, we assume the same rigidity dependence of $D_1$ and $D_2$, which only influences the slope above the softening.
Note that even in its most general form, our model contains four parameters to reproduce the two spectral breaks ($D_2$, $\delta_2$, $H_2$, $u_2$) which is the same amount of free parameters needed to explain it with an ad-hoc break and a maximum rigidity cutoff at sources ($3$ for the break and one for the cutoff). Our model does not increase the amount of parameters, but solely gives them a new meaning.

We solve the spatial part of Eq.~\ref{eq:slab} analytically and the remaining momentum-dependent equations numerically, as detailed in the Supplemental Material. To constrain our model parameters, we perform a global MINUIT \cite{james+75} fit to AMS-02 measurements of H, He, B, Be, C, N, O, Ne, Mg, Si, S, and Fe fluxes, as well as the B/C, B/O, Be/C, Be/O, C/O and He/H ratios \cite{ams21}. To capture the high-energy spectral softening, we include DAMPE measurements for H, He, C, O, and Fe \cite{dampe25}. 

Following previous studies \cite{evoli+19b,schroer+21b}, we assume all primary species (except H and He) are injected with an identical spectral index $\gamma$. We leave the outflow speeds ($u_1, u_2$), the diffusion normalization, slope, the distances ($D_0, \delta$, $H_1$, $D_2/(H_2-H_1)$), and the primary injection normalizations and slope $\gamma$ as free parameters.
The global fit yields a reduced $\chi^2\lesssim 1$ for the majority of the AMS-02 data. The notable exceptions are heavier nuclei (Si, Mg, and Fe) with $\chi^2\sim 1.4-2.4$, which were included primarily to self-consistently anchor the normalizations of the full spallation network. 
Because their fragmentation feedback on oxygen and lighter elements is small, their slightly harder spectra do not impact the derived transport parameters.
The DAMPE data were mainly used to fit the softening and determine $D_2/(H_2-H_1)$. 

\section{Results and Discussion}
\label{sec:results}

\begin{figure}
    \centering
    \includegraphics[width=\linewidth]{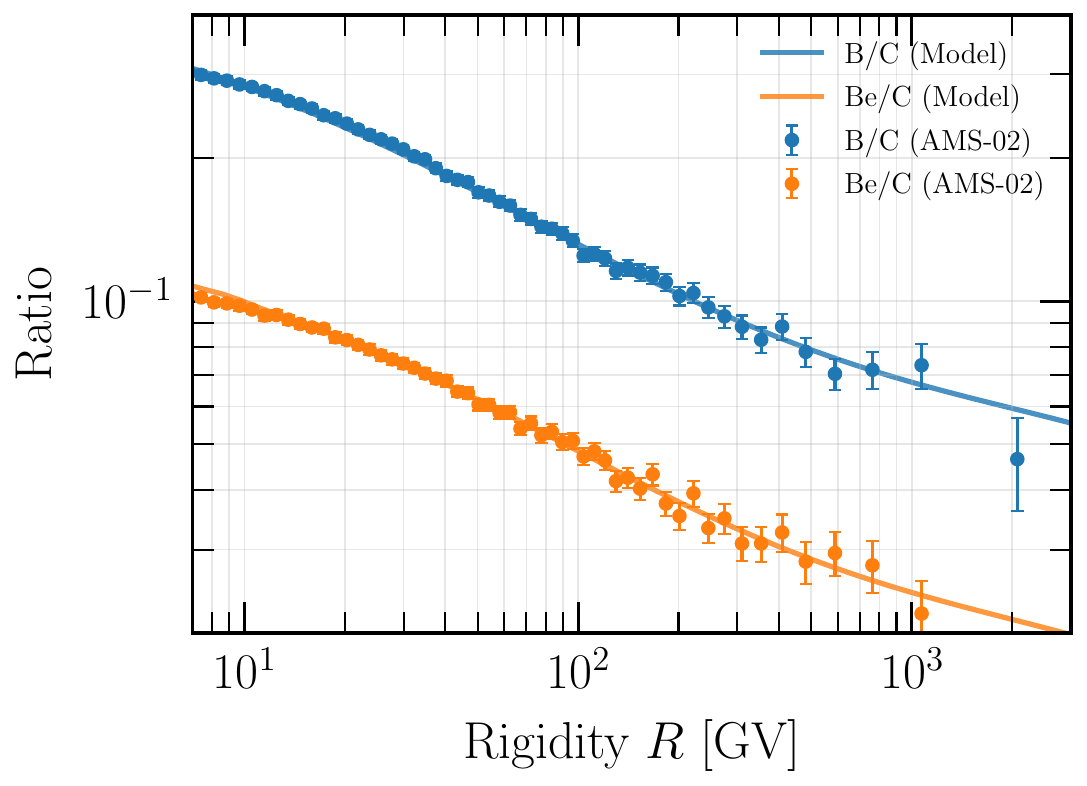}
    \caption{Boron-over-Carbon and Beryllium-over-Carbon ratios as a function of rigidity compared to AMS-02 data. Error bars represent statistical and systematic uncertainties summed in quadrature.}
    \label{fig:BC}
\end{figure}

\begin{figure*}[t]
    \centering
    \includegraphics[width=\linewidth]{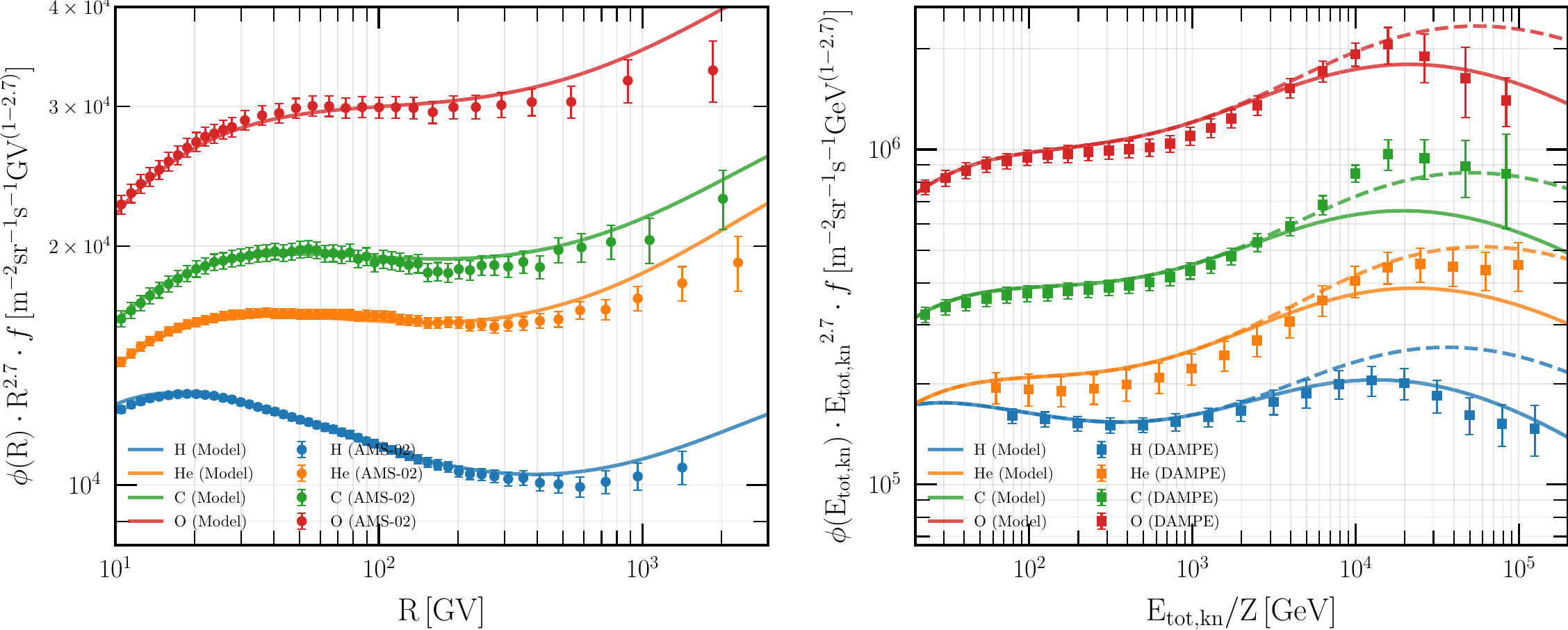}
    \caption{Left panel: Fluxes of H, He, C and O as a function of rigidity compared to AMS-02. Right panel: Same for DAMPE data as a function of total kinetic energy, divided by charge to align rigidity features. The model predictions and data of the H, He, C and O fluxes are multiplied with $f=1$, $6.3$, $233$ and $333$ respectively for illustration purposes in the left panel ($15$, $4.2$, $233$, $333$ in the right panel). Dashed lines indicate the effect of moving the softening to higher rigidities by dividing $D_2/(H_2-H_1)$ by a factor of two. }
    \label{fig:primaries}
\end{figure*}

As shown in Fig.~\ref{fig:BC}, our model successfully reproduces the B/C and Be/C spectral hardening without invoking an \textit{ad hoc} diffusion break. This feature emerges naturally from the macroscopic outflow. 
As demonstrated by the limits in Section~\ref{sec:model}, the accumulated grammage shifts from a rigidity-dependent regime ($\chi \propto 1/D$) to an advection-limited, rigidity-independent regime ($\chi \propto 1/u_2$), hardening the secondary-to-primary ratios.
The best-fit transport parameters are very similar to the usual halo model $u_1= 9\,$km/s, $D_0= 6.6\times 10^{28}$cm$^2$/s and $H_1= 5\,$kpc, with the only exception being $\delta= 0.7$, due to the slightly different rigidity dependence of our grammage.

Beyond secondary-to-primary ratios, the model successfully reproduces the high-energy features in the primary H, He, C, and O fluxes, capturing both the $300\,$GV hardening and the subsequent softening near $10\,$TV, as shown in Fig.~\ref{fig:primaries} (solid lines). Note that AMS-02 measurements appear to harden slightly less than their DAMPE equivalents for nuclei heavier than H.
Despite this trend, all fits achieved a good $\chi^2$, except for the DAMPE C flux.
This mild tension is primarily driven by the steep rise in the final three data points immediately preceding the softening. 
Hence, nuclei without such a rise (H, O) prefer a break/cutoff at lower rigidity while He and C are better fit with a break at larger rigidity.
This is illustrated by the dashed lines in Fig.~\ref{fig:primaries}, where we varied $D_2/(H_2-H_1)$ by a factor of two to mimic a larger rigidity break. While C and He fit slightly better, the fit of H and O is considerably worse.
Since in usual interpretations, such as a maximum rigidity cutoff or a transport effect, these features should be at the same rigidity for different species, it is difficult to interpret these findings.
Given how mild the tension is, we refrain here from speculating about a possible origin before the experimental uncertainties are smaller.

While the simultaneous reproduction of both spectral breaks is highly encouraging, the physical implications for each feature differ in their robustness. The hardening is a direct consequence of the eROSITA outflow. In contrast, the softening requires diffusion in the bubble which is more speculative.
In the presented fits, $D_2(10\,{\rm TV})/(H_2-H_1)= 7.2\times 10^{28}\,$cm$^2$/s/kpc fixes the softening at $10\,$TV. 
Taking the physical size of the bubble $H_2\sim 14\,$kpc would require a diffusion coefficient $D_2$ that is a factor of $\sim 13$ smaller than in the Galactic halo.
Such a smaller diffusion coefficient might be plausible given the uncertainty on the turbulence inside these outflows.
However, as we stressed above, an interpretation as a source-intrinsic effect, such as a maximum acceleration rigidity, is equally likely and strongly degenerate with the presented phenomenon. 

The distinct advantage of this advective scenario is that the derived transport parameters can be tested directly against independent multi-wavelength observations. Our global fit yields an inner halo boundary of $H_1 = 5\,$kpc and a constant outflow speed of $u_2 = 360\,$km/s. 
Remarkably, these values, derived purely from local CR flux data, align exceptionally well with the kinematic constraints on the eROSITA bubbles. The fitted boundary $H_1$ is consistent with recent models that place the closest edge of the bubbles at roughly $4\,$kpc from the Sun \cite{liu+24}. Furthermore, the required outflow speed $u_2$ falls right within the $200-1000\,$km/s range inferred from UV, radio and X-ray measurements \cite{fox+15, bordoloi+17, sarkar24,diteodoro+18,lockman+20,mccluregriffiths+13}. This agreement provides compelling evidence that macroscopic Galactic outflows, rather than purely local microphysical turbulence variations, may govern the high-rigidity transport of Galactic CRs.

\subsection{Justification of 1D Geometry and Model Uncertainties}
\label{sec:estimate}

\begin{figure}
    \centering
    \includegraphics[width=0.8\linewidth]{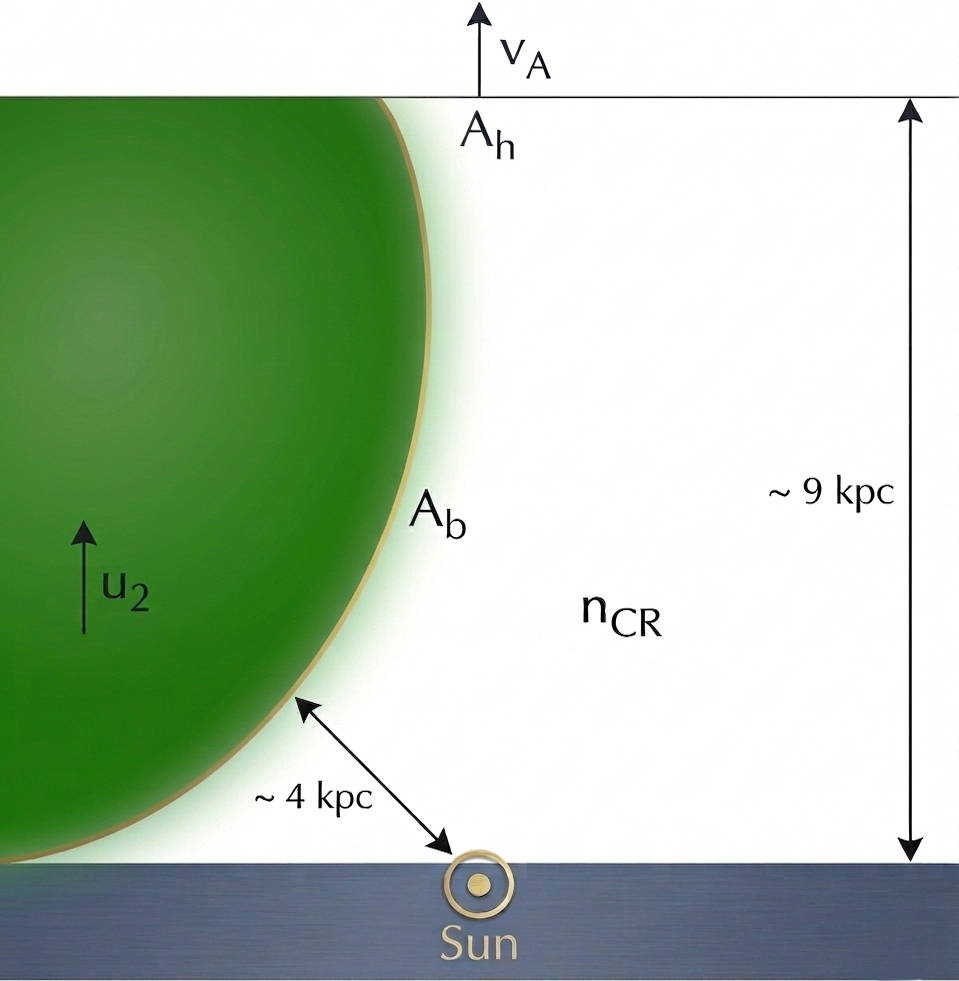}
    \caption{Sketch illustrating the escape of CRs from the Galaxy. The propagation volume is depicted as a cylinder around the Sun with height $9\,$kpc, following self-generated halo models \cite{chernyshov+22,chernyshov+24} at $1\,$TV. CRs can escape along two surfaces: The intersection $A_b$ of the bubble with the propagation volume with $u_2$ and the top surface of the halo $A_h$ with $v_A$.}
    \label{fig:sketch}
\end{figure}

Here, we estimate the importance of competing escape mechanisms, along the $z$ axis. The common mechanism to produce a limited Galactic propagation volume is self-generation \cite{evoli+18b,chernyshov+22,chernyshov+24}. In this approach, the halo size is energy dependent and corresponds to the distance at which damping becomes inefficient and particles are advected with the Alfvén speed. Inside the halo, the spatial gradients of the CR density become very flat at large energies and CR escape with a finite Alfvén speed creates a bottleneck in their escape \cite{chernyshov+22}. This picture is qualitatively similar to the 1D model we adopted, but the distance and velocity have different physical interpretations.

Due to the weak spatial variation, high-energy CRs fill their propagation volume almost uniformly, before escaping at the halo surface $A_h$ with the Alfvén speed $v_A$.
Modeling the propagation volume $V_G$ as a leaky box cylinder of radius and height $\sim 9\,$kpc (the self-generated halo size at $1\,$TV \cite{chernyshov+22,chernyshov+24}), the volume-averaged escape rate along the halo is given by $\Gamma_h\approx \frac{v_A A_h}{V_G}$. On the other hand, the propagation volume necessarily intersects with the eROSITA bubbles. 
Assuming isotropic diffusion, CRs can pass through this intersection and be advected away with the outflow $u_2$. Hence, the escape rate along the bubble is $\Gamma_b\approx \frac{u_2 A_b}{V_G}$.
The setup is illustrated in Fig.~\ref{fig:sketch}.
In a leaky box model, one would define an effective escape rate which is the sum of the two individual escape processes.
Hence, escape along the bubble dominates as long as $u_2> \frac{A_h v_A}{A_b}\approx 1.1\,v_A$, where we have used the morphology from Ref.~\cite{liu+24} to estimate $A_b/A_h$.
Note that the ratio $u_2/H_1$ is a proxy for this effective escape rate. Hence, our parameters represent an average distance and outflow speed necessary to explain AMS-02 data.

The average speed and distance being close to the observed outflow properties implies that the eROSITA outflows present at least a competitive mechanism of particle escape.
In order to achieve these velocities in self-confinement models, the background gas density decreases exponentially as a function of $z$ down to $n(z=9\,\text{kpc})\sim 2\times 10^{-5}\,$cm$^{-3}$ while keeping the temperature and the magnetic field constant up to $9\,$kpc distances \cite{chernyshov+22,chernyshov+24}.
Given the large uncertainties on these spatial profiles, it is highly plausible that the global escape rate is entirely driven by the eROSITA bubbles instead.
Once we take into account a vertical gradient of the CR density inside the halo, escape along the halo would decrease with respect to the bubble, decreasing its relative importance.

Despite good agreement with multiwavelength observations, our fit parameters suffer from the usual uncertainties in CR physics. The exact halo size is dependent on the cross sections and available Be data, where new data from the AMS-02 \cite{ams21} and HELIX \cite{helix24} experiment should soon reduce the uncertainties.
Furthermore, the outflow speed is determined by the plateau in secondary-to-primary ratios which might have a contribution from an energy-independent source grammage.
We tested a posteriori that such a source grammage of $\sim 0.4\,$g/cm$^2$ \cite{ambrosone+25,aloisio+13} would change $u_2$ to $\sim 600\,$km/s which is still compatible with multiwavelength observations.

\section{Conclusions}
\label{sec:conclusions}

Measurements of CR fluxes have revealed several unexpected spectral features, such as a hardening at $300\,$GV and a softening of several primary fluxes around $10\,$TV \cite{ams21,pamela11, dampe22,calet22,dampe21,dampe25}.
The simultaneous occurrence of the hardening in secondary-to-primary ratios suggests a link to a transport phenomenon and is often modeled with an ad-hoc break in the diffusion coefficient \cite{genolini+17,evoli+19b,schroer+21b,boschini+20,dimauro+24,delatorreluque+22}.
Although there are many possible origins for such a break related to microphysical turbulence and inhomogeneous scattering properties \cite{chernyshov+22,chernyshov+24,evoli+18b,blasi+12,ewart+26,recchia+24,tomassetti12,cowsik+14,cowsik+16,kempski+22}, we propose an alternative scenario, relating the hardening to observed, macroscopic outflows from the Galaxy.

By identifying the eROSITA bubbles as an advective boundary for local CRs, our phenomenological model naturally reproduces the hardening without introducing any break in the diffusion coefficient. 
Our best-fit parameters $H_1=5\,$kpc and $u_2=360\,$km/s represent the measurable distance and velocity of the outflow.
These values are in good agreement with
X-ray, radio and UV observations indicating $H_1\sim 4\,$kpc \cite{liu+24} and $u_2\sim 200-1000\,$km/s \cite{predehl+20,fox+15,bordoloi+17,sarkar24,ashley+20,lockman+20,mccluregriffiths+13,diteodoro+18}.
In principle, CR diffusion in these outflows could even reproduce the softening. However, this entirely depends on the unknown scattering properties of CRs in these environments and remains more speculative.

While our 1D model means these parameters should be interpreted as effective quantities, the approximation is robust.
Without an outflow limiting the diffusive halo, the leading theories suggest that the halo is self-generated \cite{chernyshov+22,chernyshov+24}.
However, the resulting propagation volume has a large overlap with the eROSITA bubbles and
as demonstrated in Sec.~\ref{sec:estimate}, escape along the bubbles becomes at least competitive with, if not more important than, escape along a self-generated halo \cite{chernyshov+22,chernyshov+24}.

The cross-disciplinary confirmation of the required transport parameters might suggest a paradigm shift: high-energy Galactic CR transport may be significantly influenced by large-scale, advective outflows driven by past Galactic Center activity.
A consequence of such a picture is that the distance to the advective boundary decreases toward the Galactic Center.
As a result, the CR spectra are expected to harden at lower rigidities, resulting in harder spectra towards the Center, an effect already observed in diffuse gamma-ray emission by Fermi-LAT \cite{fermi16,yang+16,peron+21}.
While this letter presents a promising step toward a unified multimessenger picture of our Galaxy, future 2D and 3D transport models, incorporating the exact morphology of the eROSITA bubbles, will be crucial to fully determine the impact of these structures on Galactic CRs.

\acknowledgements{
The author is grateful to D. Caprioli, P. Blasi, C. Evoli, V. Vecchiotti, M. Mukhopadhyay, A. Kravtsov, H. Chen and E. Simon for useful discussions.
This work of B.S. was partially supported by NASA grant 80NSSC24K0173, NSF grants AST- 2308021 and AST-2510951.
}

\appendix

\section{Solution of the Spatial Part}
\label{app:spatial}
In order to calculate the CR fluxes in the Galaxy, we start with Eq.~1 in the manuscript and solve the spatial part analytically.
Following the derivation of the weighted slab model, we calculate first the solution for the unstable nuclei, dividing the halo into two zones.
In the outer region, we solve the spatial part assuming a free escape boundary condition at $z=H_2$, leading to the same solution as in the modified weighted slab approach:
\begin{equation}
    I_{2,a}(z,p) = I_{a,H_1} \frac{e^{\beta_+ (z-H_1)}-e^{\beta_- (z-H_1)-(\beta_--\beta_+)(H_2-H_1)}}{1-e^{(\beta_+-\beta_-)(H_2-H_1)}}
\end{equation}
where $I_{a,H_1}$ is the flux at $z=H_1$ and $\beta_\pm = \frac{u_2}{2 D_2}\left(1\pm \sqrt{1+\frac{4 D_2}{u_2^2\tau_{\rm d,a} }}\right) =\frac{u_2}{2 D_2}(1\pm \Delta)$.
In order to solve the equation for the inner region, we use the continuity of the density at $H_1$.
With these assumptions, the solution in the inner zone is given by
\begin{align}
    I_{1,a}(z,p) = & I_{a,H_1} \frac{e^{\alpha_+z}-e^{\alpha_-z}}{e^{\alpha_+H_1}-e^{\alpha_-H_1}} \nonumber\\
    & + I_{a,0} \frac{e^{\alpha_+ H_1 + \alpha_- z }-e^{\alpha_- H_1 + \alpha_+ z}}{e^{\alpha_+H_1}-e^{\alpha_-H_1}}
\end{align}
where $\alpha_\pm$ is defined analogous to $\beta_\pm$ with the velocity and diffusion coefficient of zone $1$ and $I_{a,0}$ is the flux at $z=0$.

Next, we integrate Eq. 1 in the manuscript from $-\epsilon$ to $\epsilon$ around $z=0$ and $H_1-\epsilon$ to $H_1+\epsilon$ around $z=H_1$, taking the limit $\epsilon \rightarrow 0$.
Due to adiabatic energy losses at the halo interface, it is not possible to simplify the equations further and to express $I_{a,H_1}$ as a function of $I_{a,0}$.
Hence, the final set of coupled differential equations is
\begin{align}
    \Lambda_1 I_{a,0} + \Lambda_2 \partial_T I_{a,0} + \Lambda_H I_{a,H_1} = Q \label{eq:disk_eq}\\
    \Lambda_3 I_{a,H_1} + \Lambda_4 \partial_T I_{a,H_1} + \Lambda_0 I_{a,0} = 0  \label{eq:halo_eq}  
\end{align}
with 
\begin{align*}
    \Lambda_2 & = \left(\frac{\mathrm{d}E}{\mathrm{d}\chi}\right)_{ion} + \left(\frac{\mathrm{d}E}{\mathrm{d}\chi}\right)_{ad}\,,\\ 
    \Lambda_1 & = \frac{1}{\chi_{cr}}+\frac{1}{\tilde{\chi}} + \frac{\mathrm{d}}{\mathrm{d}E} \left( \Lambda_2 \right)\,,\\
    \Lambda_H & = -\frac{2 D_1}{\mu v} \frac{\alpha_+-\alpha_-}{e^{\alpha_+H_1}-e^{\alpha_-H_1}}\,,\\
    Q & = \frac{2h A p^2 q_{0,a}(p)}{\mu v}\,,\\
    \Lambda_0 & = -D_1 \frac{\alpha_+-\alpha_-}{e^{-\alpha_- H_1}-e^{-\alpha_+ H_1}}\,,\\
    \Lambda_3 & = D_1 \frac{\alpha_+ e^{\alpha_+ H_1} - \alpha_-  e^{\alpha_- H_1}}{e^{\alpha_+ H_1} - e^{\alpha_- H_1}} \\ & - D_2 \frac{\beta_- - \beta_+ e^{(\beta_- - \beta_+)(H_2 - H_1)}}{1 - e^{(\beta_- - \beta_+)(H_2 - H_1)}} + \tilde{u}\,,\\
    \Lambda_4 & =  \frac{u_1-u_2}{3}p \frac{\mathrm{d}T}{\mathrm{d}p}\,,\\
    \frac{1}{\tilde{\chi}} & = \frac{2D_1}{\mu v} \left( \frac{u_1}{D_1} +\frac{\alpha_+ e^{\alpha_-H_1}-\alpha_- e^{\alpha_+H_1}}{e^{\alpha_+H_1}-e^{\alpha_-H_1}}   \right)\,,
\end{align*}
$\tilde{u}= \frac{2}{3}(u_2-u_1)$ and the critical grammage $\chi_{cr}= \frac{m}{\sigma_a}$.
Note that the ionization $\left(\frac{\mathrm{d}E}{\mathrm{d}\chi}\right)_{ion}$ and adiabatic energy losses $\left(\frac{\mathrm{d}E}{\mathrm{d}\chi}\right)_{ad}$ are equivalent to the terms in the manuscript times $\frac{\mathrm{d}T}{\mathrm{d}p}/(\mu v)$, where $T$ is the kinetic energy per nucleon.

This solution is used for stable ($\Delta \rightarrow 1$) and unstable nuclei. There is one special case that requires further calculations, which is the one of stable nuclei that are produced by the decay of an unstable parent nucleus. In this case, we use the variation of constants method to solve the spatial part, similar to what was done in Ref.~\cite{evoli+19b}.
As a result, we get the following modifications to the default stable nuclei equation:
\begin{align}
    \Lambda_1 I_{a,0} + \Lambda_2 \partial_T I_{a,0} + \Lambda_H I_{a,H_1} = Q + Q_{0, \rm unst} \nonumber\\
    \Lambda_3 I_{a,H_1} + \Lambda_4 \partial_T I_{a,H_1} + \Lambda_0 I_{a,0} = Q_{H_1, \rm unst} \,. \label{eq:final_eqs}  
\end{align}

The source terms are given as follows:\\
\begin{align}
    & Q_{H_1, \rm unst}  =  \nonumber\\    & \left(\Delta \coth\left( \frac{\beta_+ \Delta (H_2-H_1)}{2}\right) - \coth\left(\frac{\beta_+ (H_2-H_1)}{2} \right)\right) \nonumber\\ & 
    \times \frac{D_2 \beta_+ I_{a', H_1}}{2}  + \frac{D_1 \alpha_+}{1- e^{-\alpha_+ H_1}}\left(q_1+q_2\right)
\end{align}
where the first part is the same as in the modified weighted slab approach \cite{evoli+19b} and the second part is coming from the solution inside the first halo
\begin{equation}
    q_1 = -\int_0^{H_1} \mathrm{d}z' \frac{1}{\alpha_+ D_1 \tau_{d, a'}} e^{-\alpha_+ z'} I_{a'}(z',p)
\end{equation}
\begin{equation}
    q_2 = \int_0^{H_1} \mathrm{d}z' \frac{1}{\alpha_+ D_1 \tau_{d, a'}} I_{a'}(z',p)
\end{equation}
with $I_{a'}(z',p)$ as the flux of the unstable isotope $a'$ as a function of $z$ and $p$.
The additional source term in the first equation is given by
\begin{equation}
    Q_{0, \rm unstable} = \frac{D_1 \alpha_+}{\mu v } \frac{q_2 + q_1 e^{\alpha_+ H_1}}{1 - e^{\alpha_+ H_1}} \,.
\end{equation}
The presented results are obtained by numerically solving the coupled Eqs.~\ref{eq:final_eqs}.

\subsection{Analytical Scaling}
In order to derive analytical scalings, we can further approximate the adiabatic energy losses by assuming $I_{a,H_1} \propto p^{-\gamma_H}$.
By replacing $\frac{p}{3} \frac{\partial I_{a,H_1}}{\partial p} \approx \frac{-\gamma_H}{3} I_{a,H_1}$, one can see that the adiabatic energy losses simply change the best-fit value of the velocities, following the same arguments as in Ref.~\cite{recchia+24}.
With these simplifications in Eq.~\ref{eq:halo_eq}, $I_{a,H_1}$ can be expressed as a function of $I_{a,0}$:
\begin{align}
    I_{a,H_1} = D_1 e^{(\alpha_-+\alpha_+)H_1} \frac{\alpha_+-\alpha_-}{e^{\alpha_+H_1}-e^{\alpha_-H_1}}
    C^{-1} I_{a,0} = \tilde{I}_{H_1,0} I_{a,0}
\end{align}
with 
\begin{align}
    C =  & D_1 \frac{\alpha_+e^{\alpha_+H_1}-\alpha_-e^{\alpha_-H_1}}{e^{\alpha_+H_1}-e^{\alpha_-H_1}}  + \frac{2+\gamma_H}{3}(u_2-u_1)\nonumber\\& - D_2 \frac{\beta_--\beta_+ e^{(\beta_--\beta_+)(H_2-H_1)}}{1-e^{(\beta_--\beta_+)(H_2-H_1)}} \,.
\end{align}
Note that in the limit $u_2 \rightarrow\infty$, the constant $C\rightarrow \infty$ and the solution reduces to the usual free escape boundary condition $I_{a,H_1}=0$.

Substituting the above solution into Eq.~\ref{eq:disk_eq}, we arrive at

\begin{equation}
    I_{a,0} \left( \frac{1}{\chi} + \frac{1}{\chi_{cr}} \right) + \Lambda_2 \partial_E I_{a,0} = Q
\end{equation}
with the analytical grammage
\begin{equation} \label{eq:app_chi}
    \chi = \left( \frac{1}{\tilde{\chi}} + \Lambda_H \tilde{I}_{H_1,0}   \right)^{-1} \,.
\end{equation}
In order to arrive at the analytical expression in the manuscript, we take the limit of stable nuclei ($\Delta\to 1$) and make the assumption of $u_2\gg u_1$ which is justified by our initial assumptions and can be verified a posteriori.

\bibliography{Total,mybib}
\end{document}